# Role of resonant inelastic x-ray scattering in high-resolution core-level spectroscopy of actinide materials

K. O. Kvashnina[1] and S. M. Butorin[2]

[1]European Synchrotron Radiation Facility, 6 rue Jules Horowitz, BP 220, 38043, Grenoble, France
[2]Department of Physics and Astronomy, Uppsala University, Box 516, S-751 20 Uppsala, Sweden

This paper provides a brief overview of applications of advanced X-ray spectroscopic techniques that take advantage of the resonant inelastic X-ray scattering (RIXS) in the hard and tender x-ray range and have recently become available for studying the electronic structure of actinides. We focus here on the high-energy-resolution X-ray absorption near edge structure (XANES) and core-to-core and core-to-valence RIXS spectroscopies at the U L and M edges of uranium compounds. The spectral features are analysed using a number of theoretical methods, such as the density functional theory in the local density approximation with an added Coulomb interaction (LDA+$U$) and full multiple scattering (FEFF) and ab-initio finite difference method near-edge structure (FDMNES) codes. In connection with presented results, the capabilities and limitations of the experimental techniques and theoretical methods are discussed.

## 1. Introduction

For many years, actinides were a fascinating topic of research but were of little practical interest. This situation has changed since nuclear power was considered as an attractive carbon-free energy source[1]. The new generation of nuclear reactors is designed to supply more electricity than previous technologies, but produce radioactive waste, which must be safely handled. This provides motivation for scientists to study the fundamental and applied properties of actinide systems in connection with their synthesis[2-3], oxidation[4], corrosion[5], reprocessing[6] and migration[7], stability at extreme conditions[8-9], environmental impact[10] and disposal of nuclear waste[11]. Obviously, the development of new and beneficial use of depleted uranium may extend the application beyond the nuclear industry[6-7, 12]. Characterization of U in different materials is essential for the future management of such substances.

Diverse properties of uranium systems come from the complexity of their electronic structure, in particular unfilled U 6d and U 5f valence shells. Theoretical modeling and predictions of changes in the electronic structure are difficult due to the complex crystal structure and the competition between different types of interactions: Coulomb and exchange, crystal field, spin-orbit coupling and orbital hybridization. Experimental observations of properties of U systems involve their own challenges including toxicity, radioactivity, and other safety issues. Thus, any method that can show the capability of discovering more information about changes in the electronic structure in different U systems is of great interest.

An element-selective probe to study the electronic structure is provided by inner-shell X-ray spectroscopy. Especially, hard X-ray spectroscopy is the ideal candidate for the actinide systems, since it does not require high-vacuum environment around the sample as, for example, in the case of soft X-ray spectroscopy. For the moment, X-ray absorption near edge spectroscopy (XANES) at the U $L_3$ edge (~17.166 keV) has been the most commonly reported in the literature[13-16]. Nevertheless, the large lifetime broadening of the *2p* level (~7.4 eV for U )[17] renders the technique less sensitive to the fine distribution of the unoccupied electronic states.

Features in XANES spectra at the U $L_3$ edge can be studied in more detail due to recent advances in employing an X-ray emission spectrometer[18-19]. The emission spectrometer is tuned to the maximum of the $L\alpha_1$ ($3d_{5/2}$ - $2p_{3/2}$) transition and the XANES spectrum is recorded by monitoring the maximum of the $L\alpha_1$ intensity as a function of the incident energy. The technique based on this setup has been named as partial fluorescence yield (PFY) or high-energy-resolution fluorescence detected (HERFD) X-ray absorption spectroscopy. Obviously, the technique takes advantage of the varying cross-section of core-to-core resonant inelastic x-ray scattering (RIXS) upon sweeping the incident photon energy throughout the edges. The advantage of such a setup is that the width of the spectral features is no longer limited by the $2p_{3/2}$ core hole lifetime but by the sharper $3d_{5/2}$ core hole width in the final state. This method was introduced by K. Hämäläinen and co-workers[18] and has been widely used for the transition metal systems (see e.g. Ref.[19]) but not so commonly for actinides[8, 20]. Here we present recent results obtained for U compounds using this setup at the U L and M edges. The information about the crystal field splitting and distribution of the U 6d and U 5f states, oxidation states of uranium during oxidation reactions are analyzed by using a number of theoretical methods. We will discuss on the capabilities and limitations of the experimental techniques and theoretical methods and areas of their application.

2. **Experimental details**

The experiments were performed at beamline ID26 of the European Synchrotron Radiation Facilities (ESRF) in Grenoble [21]. The incident energy was selected using the <311> reflection from a double Si crystal monochromator during the measurements at the U $L_3$, $M_{2,3}$ edge and using the <111> reflection during experiments at the U $M_{4,5}$ edges. Rejection of higher harmonics was achieved by three Si mirrors working under total reflection for the measurements at the U M edges and three Cr/Pd mirrors for the U $L_3$ edge measurements. The beam size was estimated to be ~ 0.3 mm vertically and 1 mm horizontally. XANES spectra were simultaneously measured in total fluorescence yield (TFY) mode using a photodiode and in HERFD mode using an X-ray emission spectrometer. The sample, analyzer crystal and photon detector (avalanche photodiode) were arranged in a vertical Rowland geometry. The core-to-valence RIXS spectra were recorded at a scattering angle of 90° in the horizontal plane using only one crystal analyzer. All HERFD-XANES measurements were performed using four Si <2 2 0> or four Ge <1 1 1> or four Ge <2 2 0> crystal analyzers in different reflections (summarized in Table 1). The difference in the energy resolution arises from the different Bragg angles of the crystal analyzer and different monochromator crystals. The intensity was normalised to the incident flux.

RIXS data are shown here as a contour map in a plane of incident and transferred photon energies, where the vertical axis represents the energy difference between the incident and emitted energies. Variations of the colour on the plot correspond to the different scattering intensities. Experiments reported here were performed at room temperature in air without any additional environment around the sample. The paths of the incident and emitted X-rays through the air during the experiment at the U M edges were minimized in order to avoid losses in intensity due absorption in the air. The polycrystalline materials $U_3O_8$, $U_4O_9$, uranyl nitrate hexahydrate $UO_2(NO_3)_2(H_2O)_6$, torbernite mineral $Cu(UO_2)_2(PO_4)_2(H_2O)_{12}$ and single crystal $UO_2$ were covered by 50 μm Kapton film and did not show any radiation damage.

The $U_4O_9$ powder was prepared by heat treatment of a mixture of $UO_2$ and $U_3O_8$ powders. The relative mass fraction of $UO_2$ and $U_3O_8$ was chosen in order to get, on average, a $UO_{2.23}$ composition. The powders were mixed carefully, placed in an air-tight closed quartz tube and underwent a heat treatment at 1050°C for 30 days. After that they were slowly cooled down to room temperature over 12 h. The powder obtained had a dark colour. The quality was checked by X-ray diffraction and contained less than 1% $U_3O_8$, assuming that the $U_4O_9$ phase in the sample had an oxygen composition that was very close to the phase stability limit in the phase diagram. The $U_3O_8$ powder was obtained via isothermal annealing at 900 K of a sintered $UO_2$ pellet in dry air. The X-ray diffraction pattern showed the expected structure. For the HERFD- XANES experiment, 10 mg of $U_4O_9$ or $U_3O_8$ were diluted in 200 mg of boron nitride and then pressed to obtain a solid pellet. To avoid any further oxidation of these two samples, after preparation each pellet was immediately put in a sealed copper sample holder with a 5 μm Kapton window and once more covered by 50 μm Kapton film.

Table 1. Reflections of the crystal analysers used in the detection of U x-ray emission lines, together with the corresponding Bragg angles and energy resolutions estimated.

| **Emission energy** | **Crystal analyzer** | **Bragg angle** | **Resolution** |
|---|---|---|---|
| HERFD at the U $L_3$ edge at $L\alpha_1$ ($2p_{3/2}$-$3d_{5/2}$) = 13.616 keV | Ge <7 7 7> | 77° | 2.2 eV |
| HERFD at the U $L_3$ edge at $L\beta_2$ ($2p_{3/2}$-$4d_{5/2}$) = 16.388 keV | Si<10 10 0> | 81° | 2.0 eV |
| HERFD at the U $L_3$ edge at $L\beta_5$ ($2p_{3/2}$-$5d_{5/2}$) = 17.063 keV | Si<10 10 0> | 71° | 2.2 eV |
| Valence band RIXS at the U $L_3$ edge, E = 17.166 keV | Ge<9 9 9> | 84° | 1.6 eV |
| HERFD at the U $M_2$ edge at ($3p_{1/2}$ – $5d_{3/2}$) = 5079 eV | Ge<3 3 1> | 71° | 1.2 eV |
| HERFD at the U $M_3$ edge at ($3p_{3/2}$ – $4d_{3/2}$) = 3522 eV | Si<2 2 0> | 67° | 1.2 eV |
| HERFD at the U $M_4$ edge at $M\beta$ ($4f_{5/2}$ -$3d_{3/2}$) = 3336 eV | Si<2 2 0> | 75° | 0.7 eV |
| Valence band RIXS at the U $M_5$ edge, E ~ 3552 eV | Si<2 2 0> | 65° | 1.1 eV |
| HERFD at the U $M_5$ edge at $M\alpha_1$ ($4f_{7/2}$ -$3d_{5/2}$) = 3171 eV | Ge<2 2 0> | 78° | 0.9 eV |

3. **Theoretical calculations**

   a. **FDMNES calculations**

The XANES spectrum at the U $L_3$ edge of $UO_2$ was simulated by the finite difference method for near-edge structure (FDMNES)[22]. The code allows the calculation of the occupied and unoccupied projected density of states (DOS) in relation to the X-ray absorption and emission processes. Simulations were performed using an atomic cluster with 4.6 Å radius for $UO_2$ using Cartesian coordinates listed in the Inorganic Crystal Structure Database (ICSD). $UO_2$ has a fluorite-type structure (Fm3m) with lattice parameter a= 0.571 nm, where each U atom is surrounded by eight O atoms. Relativistic self-consistent field calculations using the Dirac-Slater approach have been performed for each atom in the considered cluster. From the superposed self-consistent atomic densities in the selected cluster, the Poisson equation was solved to obtain the Coulomb potential. The energy-dependent exchange correlation potential was evaluated using the local-density approximation. The exchange-correlation potential was constructed using both the real Hedin-Lundquist and the Von Barth formulations[20].

### b. LDA + *U* calculations

The first-principles calculations of $UO_2$ and $UO_2(NO_3)_2 \cdot 6H_2O$ were performed using the linearized muffin-tin orbitals method LMTO-ASA[23] in the atomic sphere approximation. The radii of muffin-tin spheres were set to R(U) = 1.73 a.u. and R(O)=1.6 a.u. Interstitial space was filled with empty spheres. Structural relaxation effects were neglected. In the calculations of $UO_2$ the lattice parameter of 0.5371 nm was used, which corresponds to the minimum of the energy of the system[24]. In order to reproduce the antiferromagnetic order observed in earlier experiments[25] a √2 x √2 x 1 supercell dimension was selected. In the calculations of $UO_2(NO_3)_2 \cdot 6H_2O$ the atomic positions and crystal structure identified by experiment from Taylor and co-workers[26] were used. The LMTO basis set contained the following states: U(7s,6p,6d,5f), O(2s,2p), N(2s,2p), H(1s). Correlation effects for 5f states of uranium were taken into account by using the LDA+*U* [27] approximation with additional Coulomb *U* parameter on the U atom. The values of the on-site Coulomb interaction parameters *U* and Hund's exchange *J* were taken to be 4.5 eV and 0.5 eV, respectively for both simulations of $UO_2$ and $UO_2(NO_3)_2 \cdot 6H_2O$ . These values are considered to be reasonable for this type of system and have been used in previous calculations[24] . The calculated value of the band gap of $UO_2$ was found to be 1.9 eV, which agrees with the experimentally determined value of 2.0 eV [25]. The partial densities of states of $UO_2$ were calculated using the 6 x 6 x 4 k-point grid in the entire Brillouin zone. The value of the U magnetic moment of U atom was estimated to be 1.91 $\mu_b$, which is also in agreement with results of previous investigations[24]. A Brillouin zone integration for $UO_2(NO_3)_2 \cdot 6H_2O$ has been performed on the 2 x 2 x 1 k-point grid. The value of the band gap denoted from the calculation was estimated to be ~3.06 eV. To the best of our knowledge, this is the first LDA +*U* calculation of $UO_2(NO_3)_2 \cdot 6H_2O$ reported. The detailed discussion of the calculated results will be published in Ref. 28.

Core-to-valence RIXS calculations at the U $L_3$ edge were performed by inserting the U 6d density of states into the Kramers-Heisenberg equation:

$$F(\Omega, \omega) = \int_\varepsilon d\varepsilon \frac{\rho(\varepsilon)\rho'(\varepsilon + \Omega - \omega)}{(\varepsilon - \omega)^2 + \frac{\Gamma_n^2}{4}} \quad (1)$$

Where ρ and ρ' are the density of occupied and unoccupied U 6d states, while $\Omega$ and $\omega$ represent the energies of the incident and scattered photons, respectively. $\Gamma_n$ represents the lifetime broadening of the U $2p_{3/2}$ state, which is 7.4 eV[17]. The validity of this approximation[29] can be evaluated by comparison between experimental and theoretical results. An elastic peak due to Thomson scattering was added to the RIXS planes to facilitate comparison with experiment.

### c. FEFF calculations

The full multiple scattering code FEFF 8.4 [30] was used to model the experimental XANES spectra of $UO_2$ and $Cu(UO_2)_2(PO_4)_2(H_2O)_{12}$ at the U $M_4$ edge. The input files are based on the crystal structure of $UO_2$ and $Cu(UO_2)_2(PO_4)_2(H_2O)_{12}$ reported in the literature[31-32].

Calculations are made for the central U atom in a cluster of 75 atoms (6.0 Å radius) for UO$_2$ and for the central U atom in a cluster of 440 atoms (10.0 Å radius) for Cu(UO$_2$)$_2$(PO$_4$)$_2$(H$_2$O)$_{12}$. Full multiple scattering (FMS) calculations are performed using a Hedin-Lundqvist self energy correction and other standard cards. The reduction of the core hole lifetime broadening is achieved by using the EXCHANGE card. More details about the FEFF code are reported elsewhere [30].

## 4. Results and Discussions

Conventional X-ray absorption spectroscopy has been successfully used in studies of U materials, especially in the hard X-ray range at the U $L_3$ edge. It is applied to probe the local electronic structure at U atoms, for the speciation and determination of the oxidation state of U in various systems. However, the large lifetime broadening (~7.4 eV[17]) of the U 2p level leaves the possibility of examining only the post-edge region of the X-ray absorption spectra[33] at the U $L_3$ edge. Some authors analyze the chemical shift of the U $L_3$ white line by taking the first derivative value, in order to determine the U valence state[34]. However, direct evidence of the shift in energy as being due to the different oxidation state is not easy to achieve. Recent advances in spectroscopy demonstrated that the spectral broadening of the absorption features can be reduced by employing an X-ray emission spectrometer.

Figure 1a shows a schematic representation of the processes which can be measured using an X-ray emission spectrometer, named as core-to-core RIXS and core-to-valence RIXS. As an example Fig. 1b contains the core-to-core RIXS profile map recorded by scanning the incident energy at different emission energies around the U $L\alpha_1$ emission line of UO$_2$. A scan at the maximum of the $L\alpha_1$ emission line is referred as HERFD-XANES spectrum. Fig. 1b reveals that no additional peaks away from the HERFD dashed line are observed in the RIXS map and the detailed analyses of the features can be performed based on the HERFD-XANES scan.

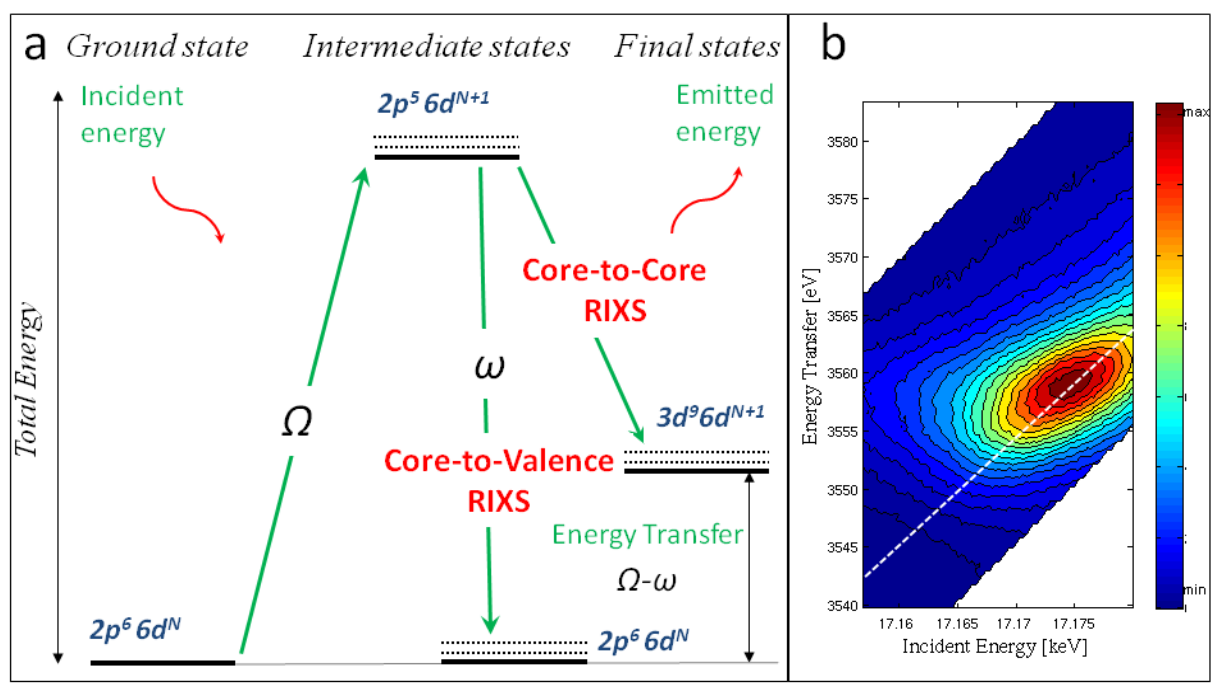

Fig. 1. a) Schematic diagram of core-to-core RIXS and core-to-valence RIXS processes. b) Core-to-core RIXS plane displayed as a contour map with axes corresponding to the incident and transferred energies around the U $L\alpha_1$ emission line for UO$_2$. A U $L_3$ HERFD-XANES

spectrum corresponds to the diagonal cut (dashed line) through the RIXS plane at the maximum of the L$\alpha_1$ emission line.

Fig. 2 shows the comparison of HERFD and conventional XANES (TFY-mode) spectra at the U L$_3$ edge of $UO_2$ and $UO_2(NO_3)_2 \cdot 6H_2O$, which are nominally tetravalent and hexavalent U systems, respectively. The broadening of the spectral features in the HERFD spectrum is determined by the lifetime broadening of the final state (3d$_{5/2}$ core hole) and the instrumental resolution (as a combination of incident and emission energy bandwidths (~2.2 eV))[35-36]. Using the formula given by Swarbrick and co-workers[36] we estimate the total spectral broadening of the HERFD spectrum at the U L$_3$ edge to be ~ 4.7 eV. Observed features in the HERFD spectra[20] at the U L$_3$ edge of U systems are much sharper compared to those in the TFY spectra (Fig. 2).

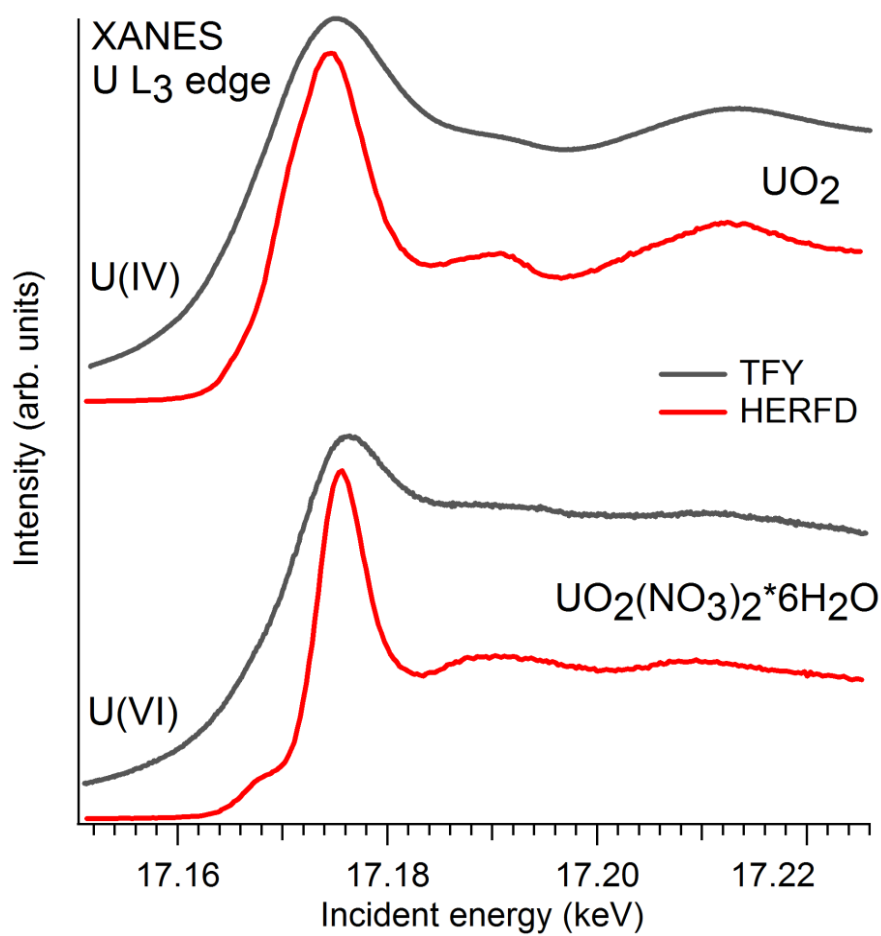

Fig. 2 U L$_3$ HERFD-XANES of $UO_2$ and $UO_2(NO_3)_2 \cdot 6H_2O$ recorded with the X-ray emission spectrometer tuned to the maximum of the L$\alpha_1$ emission line at 13.616 KeV. The spectra are compared with total fluorescence yield (TFY) curves recorded using a photodiode.

The L$\alpha_1$ X-ray emission line has the highest transition probability compared with other emission lines, thus providing a strong signal during the data collection. However, the spectral broadening can be reduced further if the X-ray emission spectrometer is tuned to an emission line with a smaller core-hole lifetime width. Fig. 3 shows the HERFD spectra of $UO_2$ at the U L$_3$ edge recorded using different emission lines: L$\alpha_1$ (*3d$_{5/2}$ - 2p$_{3/2}$*), L$\beta_2$ (*4d$_{5/2}$ - 2p$_{3/2}$*) and L$\beta_5$ (*5d$_{5/2}$ - 2p$_{3/2}$*). The lifetime broadening of the HERFD-XANES spectrum recorded at the L$\beta_2$ emission line is similar to that at the L$\alpha_1$ line. The 5d core hole (in case of measurements at the L$\beta_5$ line) produces a smaller lifetime broadening (~1eV). We estimate the total spectral broadening of the HERFD spectrum at the U L$_3$ edge collected at the maximum of the L$\beta_5$ emission line to be ~ 2.4 eV. Fig. 3 shows the experimental results for HERFD-XANES collected at different emission lines which are compared to the results of the theoretical simulation using the FDMNES code. Following the dipole selection rule at the U L$_3$ edge, electrons are excited from the 2p$_{3/2}$ core level to the unoccupied 6d states. The theoretical spectrum illustrates the distribution of the unoccupied projected U 6d density of states (DOS) of UO$_2$. Comparison between theoretical and experimental HERFD spectra collected using the L$\alpha_1$ emission line shows that spectral features A and B are not experimentally resolved. The resolution in terms of A and B distinction in the U L$_3$ XANES spectrum was improved only due to recording HERFD-XANES at the L$\beta_5$ emission line. The analysis shows that the A and B features are a result of the crystal-field splitting of the U 6d states between e$_g$ and t$_{2g}$ orbitals.

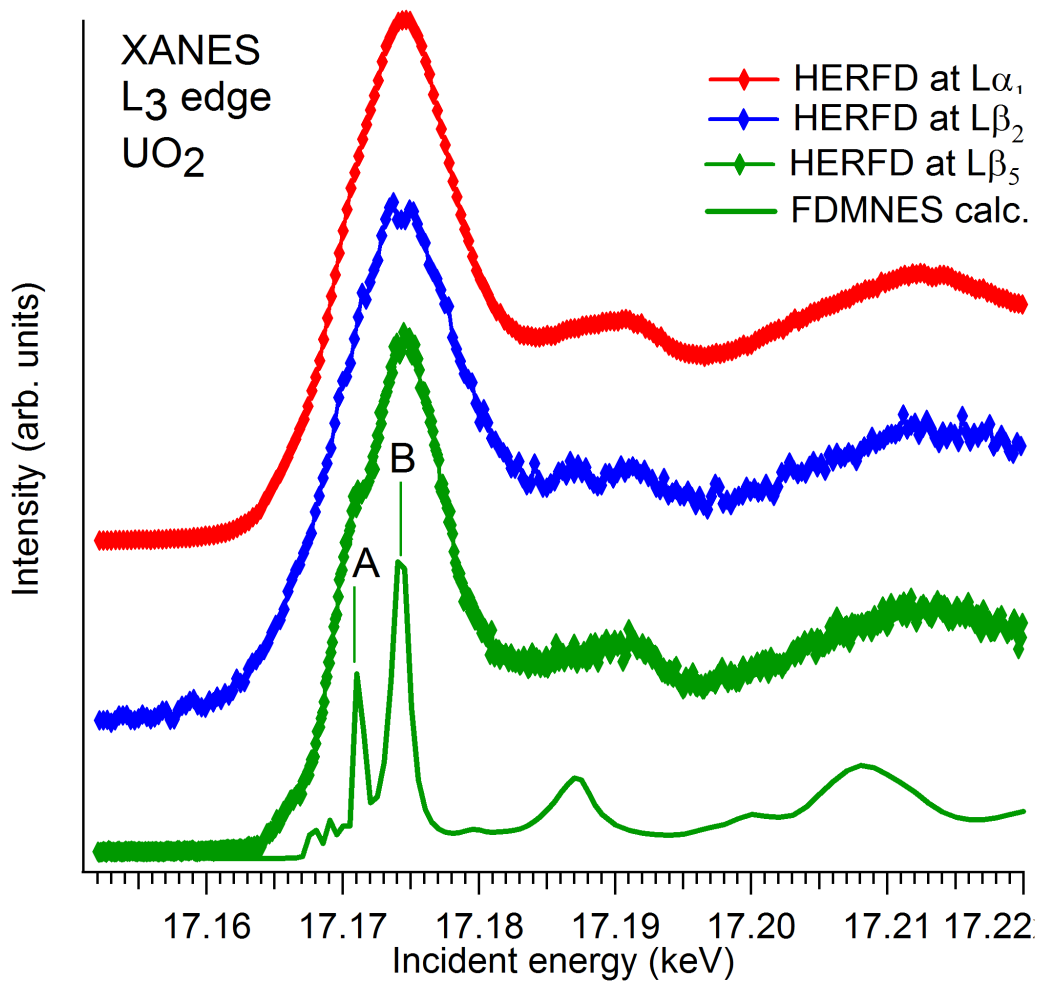

Fig. 3 HERFD-XANES at the U L$_3$ edge of UO$_2$ recorded with X-ray emission spectrometer set to different X-ray emission lines: L$\alpha_1$ (red), L$\beta_2$ (blue) and L$\beta_5$ (green). Results of the FDMNES calculations are also shown.

The way to probe the crystal field splitting of the U 6d states in the ground state of the system is to perform core-to-valence RIXS measurements (see Fig. 1). The intermediate state in core-to-core and core-to-valence RIXS is the same and has a core hole making the technique element selective. The final state may or may not contain a core hole depending on the energy transfer that is chosen in the experiment. The energy transfer in core-to-core RIXS near the L$\alpha_1$ emission line is very large (~3552 eV). In core-to-valence RIXS the decay directly involves valence electrons. The energy transfer is of only a few electron volts and no core hole is present in the final state (see Fig. 1). In this case, the crystal-field splitting of the U 6d states is not distorted by the interaction with existing core hole.

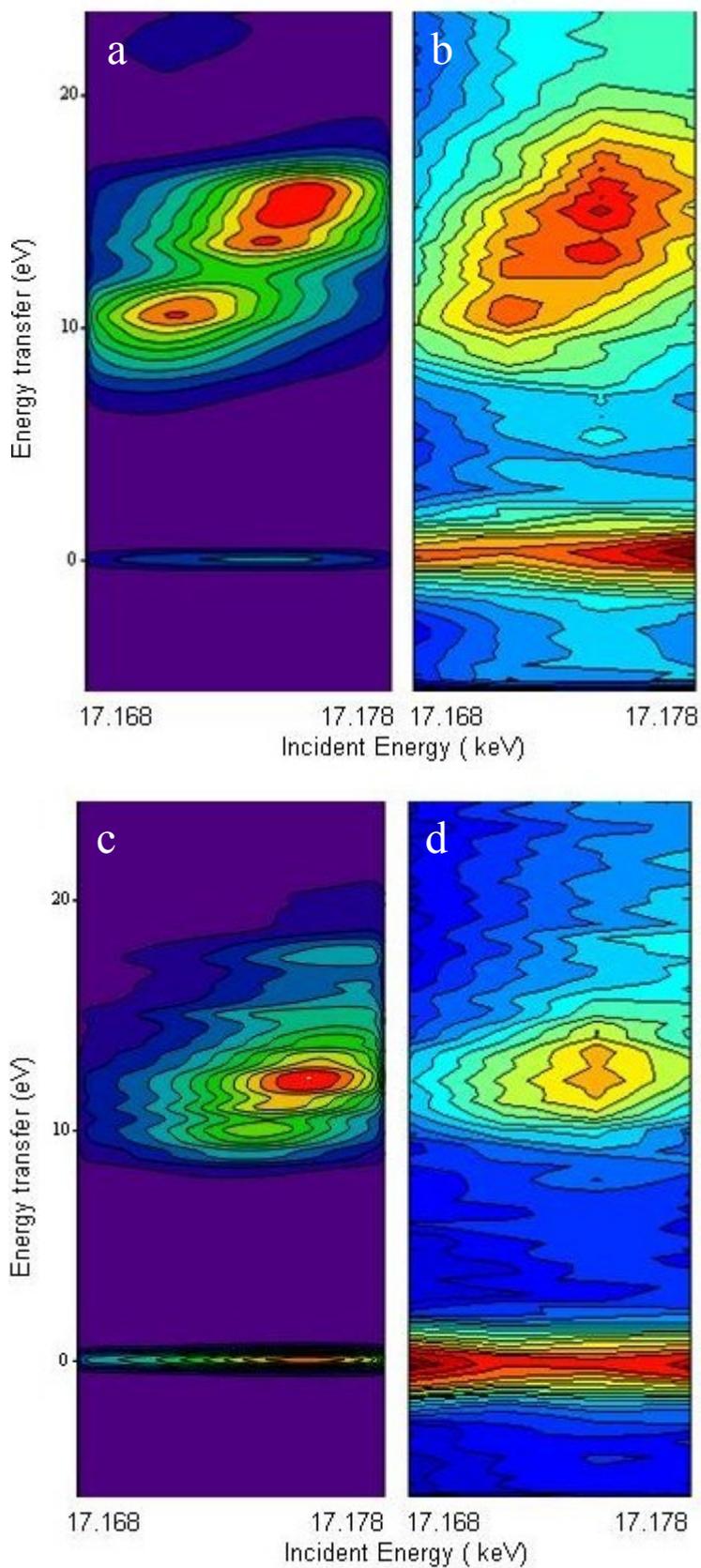

Fig. 4. Experimental (b,d) and theoretical (a,c) core-to-valence RIXS intensities displayed as contour maps with axes corresponding to incident and transferred energies at the U $L_3$

absorption edge of UO$_2$ (a,b) and UO$_2$(NO$_3$)$_2 \cdot$6H$_2$O (c,d). Variations of the colour in the plot relate to the different scattering intensities.

Fig. 4 shows the experimental (b,d) and theoretical (a,c) core-to-valence RIXS profiles of UO$_2$ (a,b) and UO$_2$(NO$_3$)$_2 \cdot$6H$_2$O (c,d). The experimental data were obtained with excitation energies around the maximum of the U L$_3$ absorption edge. Two types of transitions are observed: elastic scattering peak at ~ 0eV and inelastic scattering features at higher energy transfer. In the core-to-valence RIXS plane, the 2p$_{3/2}$ core hole is still present along the horizontal scale of incident energy (as in the core-to-core RIXS plane) but no core hole is present in the final state (vertical scale). The core-to-valence RIXS calculations at the U L$_3$ edge were performed by inserting the U 6d DOS into the Kramers-Heisenberg equation (1).

The idea to relate the RIXS spectral features to the DOS was introduced by Jimenez-Mier and co-authors[29]. The RIXS process at the U L$_3$ edge is identified as the convolution of the occupied and unoccupied U 6d DOS. Here we used the projected DOS obtained in LDA+$U$ calculations (Fig. 5). The Fermi level at 0 eV in Fig. 5 divides the occupied and unoccupied projected DOS. Analysis of equation (1) shows that the energy difference between the occupied and unoccupied U 6d states will correspond to the energy transfer values for the observed RIXS transitions. Theoretical and experimental RIXS maps for UO$_2$ shown in Fig. 4 demonstrate the splitting of the RIXS features in the valence band into those with energy transfer of ~10 eV and ~15 eV. These features arise from the peaks A' and B' in the projected U 6d DOS of UO$_2$ (Fig. 5). These are similar to features A and B obtained in the FDMNES calculations in Fig. 3 that are a result of the crystal field splitting of the U 6d states. No splitting is detected in the RIXS map for UO$_2$(NO$_3$)$_2 \cdot$6H$_2$O but the observed transitions at ~12 eV energy transfer in the experimental data are identical to those in the theoretical calculations.

We find a reasonably good agreement between theoretical and experimental core-to-valence RIXS results at the U L$_3$ edge for both compounds, with respect to feature shapes, positions and relative intensities. The disadvantage of the LDA+$U$ approach is that $U$ comes as a parameter and is usually chosen to fit experimental data, for example, the value of the band gap and magnetic moment. Nevertheless, in addition to reasonable considerations for the choice of the $U$ value there are as well some theoretical methods to estimate it self-consistently. In any case, theoretical simulations of the RIXS data can be performed by utilizing any of the theoretical codes that describe correctly the empty and occupied projected DOS.

An inspection of the projected DOS obtained by the LDA + $U$ calculations for UO$_2$ (Fig. 5) and UO$_2$(NO$_3$)$_2 \cdot$6H$_2$O reveals that the main contribution to the DOS above the Fermi level comes from the U 5f states with some admixture of ligand 2p states. The hybridization between U 5f and U 6d states for both compounds is observed in theoretical calculations. The significant U 5f weight is much closer to the Fermi level than that of the U 6d states, therefore probing the U 5f states should be more important in terms of characterization of the electronic structure of U. Knowledge about the contribution of U 5f states can be obtained by performing the RIXS measurements at the U M$_{4,5}$ edges (3.5-3.7 keV) that characterize the U 3d → 5f electronic transitions.

The experimental core-to-core RIXS maps for incident photon energies at the U L$_3$ and the U M$_4$ edges of UO$_2$ are shown together in Fig. 6. The dashed lines through the RIXS planes indicate the corresponding HERFD spectra. Similarly to the core-to-core RIXS data around the U Lα$_1$ emission line, no additional peaks away from the HERFD dashed line are observed in RIXS plane at the U Mβ emission line. Therefore, an analysis of the observed features can be based on the HERFD-XANES data.

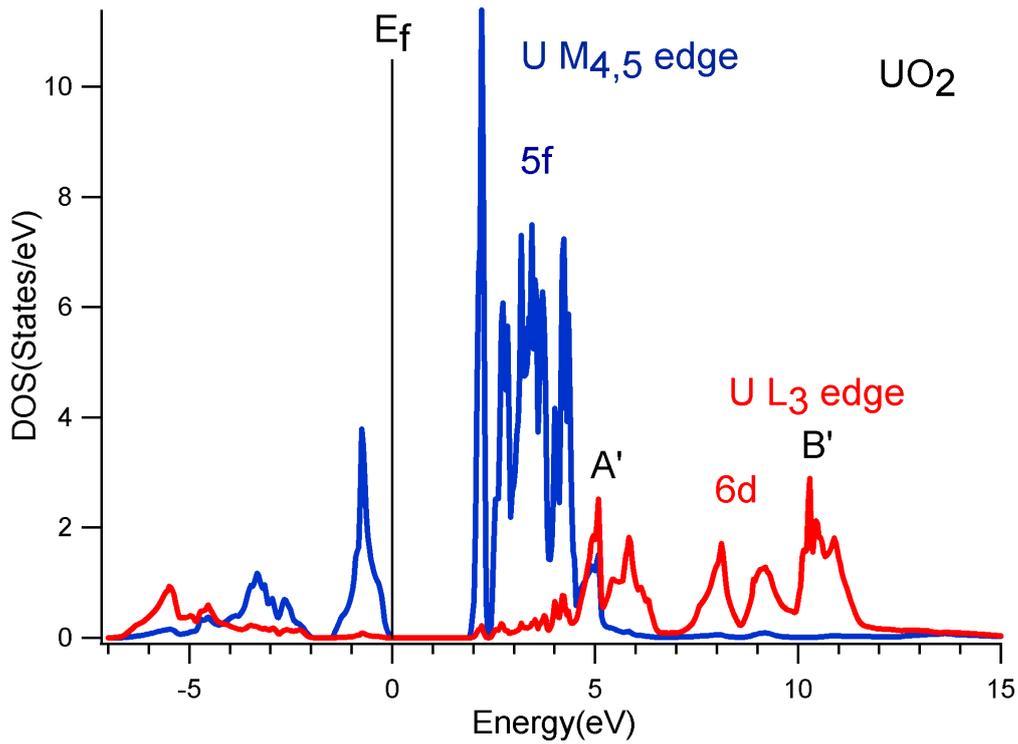

Fig. 5. Projected U5f (blue) and U6d (red) density of states of $UO_2$ as obtained in LDA+ $U$ calculations in Ref. 28.

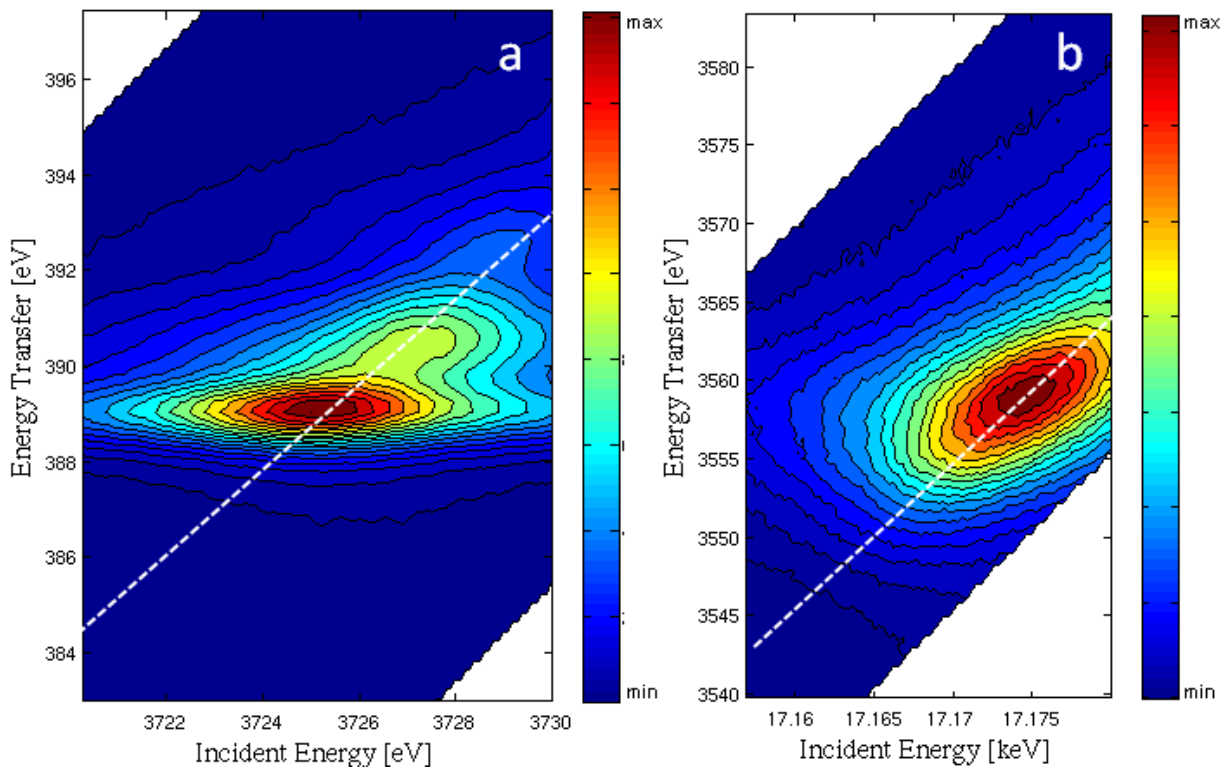

Fig. 6 Experimental core-to-core RIXS intensities displayed as contour maps with axes corresponding to incident and transferred energies at the U $M_4$ (a) and the U $L_3$ (b) absorption edges of $UO_2$. Variations of the colour in the plot relate to the different scattering intensities. A HERFD spectrum corresponds to a diagonal cut (dashed line) through the RIXS plane at the maximum of the $L\alpha_1$ and $M\beta$ emission lines.

Fig. 7 shows XANES spectra at the U $M_4$ edge of $UO_2$ collected in the HERFD and TFY modes. The shape and the position of the main line in the TFY spectrum of $UO_2$ is in agreement with previously reported results. The HERFD-XANES spectrum of $UO_2$ shows a sharp peak at ~3725 eV, due to the transitions from the $3d_{3/2}$ core level to the unoccupied $5f_{5/2}$ level, and a shoulder at higher energy. The difference of the HERFD-XANES spectrum of $Cu(UO_2)_2(PO_4)_2(H_2O)_{12}$ from that of $UO_2$ is in splitting of the main edge transitions into two peaks at ~3727 eV and ~3729 eV. Several research groups[14, 37,38] previously reported a shoulder on the high-energy side of the main line in the conventional XANES at the U $M_4$ edge for other hexavalent U systems. The shoulder was associated with the characteristic signature of uranyl groups ($UO_2^{2+}$) in those U systems[37]. We confirmed the assignment by performing multiple-scattering calculations using FEFF8.4 where the input file was based on the crystal structure of $Cu(UO_2)_2(PO_4)_2(H_2O)_{12}$ reported by Locock and Burns[31].

Fig. 7 shows the calculated XANES spectrum of $Cu(UO_2)_2(PO_4)_2(H_2O)_{12}$ at the U $M_4$ edge for comparison with experimental data. The splitting of the main edge transitions is well reproduced and is related to the distribution of the U 5f states. There is also a very good correspondence between theoretical and experimental HERFD-XANES spectra of $UO_2$ with respect to the shoulder of the main U $M_4$ line. However, we were not able to reproduce the spectral feature at ~3733 eV in the HERFD-XANES spectrum of $Cu(UO_2)_2(PO_4)_2(H_2O)_{12}$. This indicates that the interaction of the 5f electrons with the core hole is important and needs to be taken into account. The short-range models, such as the Anderson impurity model, explicitly treat the interaction with the core hole but the approach is rather parametric and the outcome depends on the values of a few physical quantities (parameters). In the Anderson impurity model, The ~3733 eV transition can be associated with ligand 2p - U 5f charge transfer excitations. The observed spectral difference between $UO_2$ and $Cu(UO_2)_2(PO_4)_2(H_2O)_{12}$ as tetravalent and hexavalent U systems suggests that HERFD-XANES is a technique with enhanced sensitivity to distinguish between the U oxidation states in various U materials.

As an example, we discuss here a direct identification of the U valence states in mixed uranium oxides $U_3O_8$ and $U_4O_9$. Mixed uranium oxides are one of the technologically and ecologically important classes of materials due to their participation in nuclear fuel cycles, long-term storage and environmental migration issues. The U oxidation states in mixed U oxides were studied before by using XANES at the U $L_3$ edge. Up to now it was often assumed that $U_3O_8$ has a mixture of U(IV) and U(VI) valence states. The question about the valence states of U in $U_4O_9$ is also not entirely clear. Most researchers claim the existence of U(IV) and U(V) valencies. However, Conradson and co-workers[34] observed small U-O distances and oxo-groups during the analysis of the post-edge absorption region of $U_4O_9$ at the U $L_3$ edge, which are characteristic for U(VI) compounds. Experimental data reported here for the U $L_3$ edge (Fig. 2) show that the chemical shift of the U main line between U(IV) and U(VI) systems is not that large (~1 eV). Therefore, even an improved spectral resolution due to reduced core-hole lifetime broadening in the HERFD experiment may not provide a clear distinction between U sites in complex systems. TFY spectra at the U $M_{4,5}$ edges of U systems (see e.g. Fig. 8) are broad as well and show a small energy shift of the main absorption line with changing U valency. Under these conditions, the detection of the U oxidation states is challenging, especially for small concentrations of species in question.

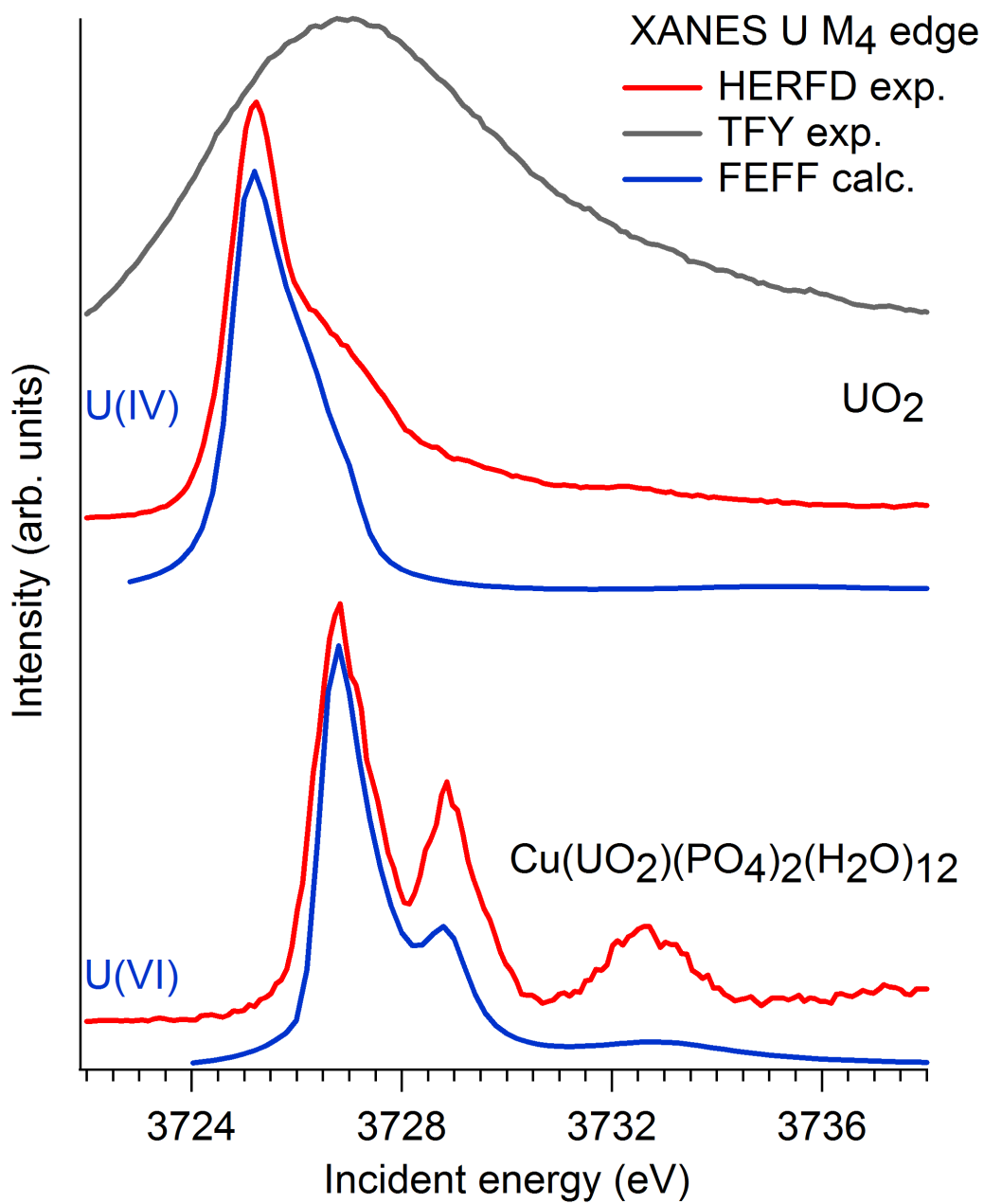

Fig. 7 High energy resolution (HERFD) U $M_4$ XANES of $UO_2$ and torbernite $Cu(UO_2)_2(PO_4)_2(H_2O)_{12}$ recorded with the X-ray emission spectrometer set to the $M\beta$ emission line at 3336 eV, and compared with multiple scattering calculations using the FEFF 8.4 code. The HERFD spectrum for $UO_2$ is compared with the total fluorescence yield (TFY) curve recorded using a photodiode.

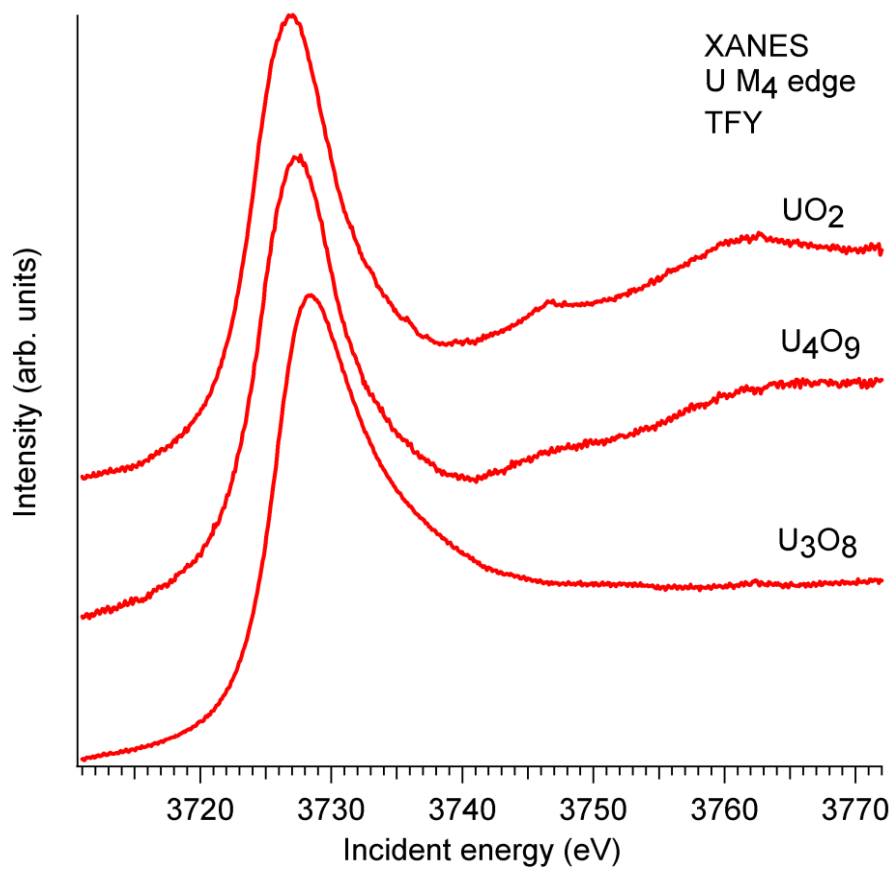

Fig. 8 XANES spectra of uranium oxides at the U $M_4$ edge measured in the total fluorescence yield (TFY) mode.

Fig. 9 shows experimental HERFD-XANES data at the U $M_4$ edge for $U_4O_9$ and $U_3O_8$ (taken from Ref. 39) compared to the tetravalent and hexavalent reference systems. The positions of the main peaks in the HERFD spectra of the reference U systems clearly reveal the energy shift of ~1.9 eV of the white line towards higher energy on going from the U(IV) to U(VI). This shift is considerably larger than what was previously observed at the U L edge.

Width and shape of the absorption lines in $U_3O_8$ and $U_4O_9$ are very different from those in $UO_2$ and torbernite. Two intense peaks in the HERFD spectrum of $U_4O_9$ are observed. The first peak at ~3725 eV is associated with the U(IV) signal. The chemical shift of the second peak is smaller than that for U(VI) and therefore can be attributed to U(V). The width of the $U_3O_8$ white line is smaller than that for $U_4O_9$ and an asymmetry at the higher energy side is notable. We assign the intense structure at ~3726 eV to a U(V) contribution. The obtained results indicate that the HERFD technique at the U M edges can be utilized as a fingerprint method in the detection of the U oxidation states in various complex systems such as soils, solutions/waters and minerals containing U ions.

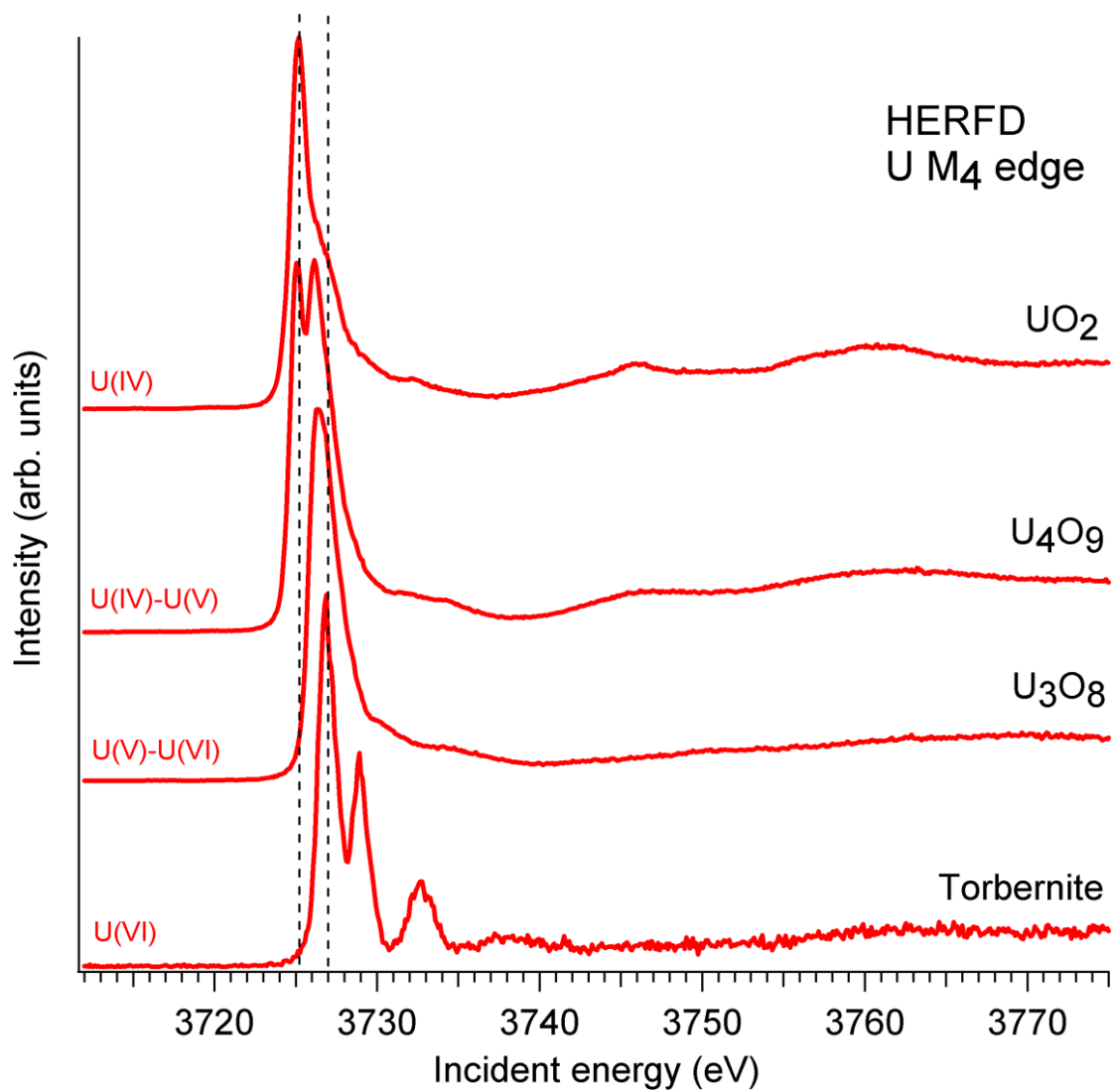

Fig. 9. High energy resolution (HERFD) U $M_4$ XANES of $UO_2$, $U_3O_8$, $U_4O_9$ and torbernite $Cu(UO_2)_2(PO_4)_2(H_2O)_{12}$ recorded with the X-ray emission spectrometer tuned to the maximum of the Mβ emission line at 3336 eV (Ref. 39).

In order to evaluate the utility of the core-to-valence RIXS method at the U M edges we performed an experiment at the U $M_5$ edge of $UO_2$ (Fig. 10). The RIXS map in Fig. 10a was measured at a number of excitation energies around the maximum of the U $M_5$ absorption line with help of five crystal-analyzers (see Table 1). The horizontal axis in Fig. 10a corresponds to the difference between the incident and emitted energies indicated as energy transfer. The vertical axis represents the incident photon energy. The position of the elastic peak which is set to 0 eV was calibrated by measuring the elastic scattering from the polyether ether ketone (PEEK) polymer. In that case the elastic peak was intense. Looking at the map in Fig. 10a, it is clear the the contribution of the elastic scattering is not the highest one. The strongest spectral intensity appears to be in the 0-2 eV energy-transfer range. The one-dimentional RIXS spectra plotted in Fig. 10b reveal a number of the RIXS features appearing in the ~0-2 eV and ~ 4.5 - 8.0 eV regions and at ~19.0 eV on the energy transfer scale.

Features in the range of ~4.5 - 8.0 eV were already observed by Butorin[38] and were assigned to the metal-to-ligand charge transfer excitations between U *5f* and O *2p* states. Later the same research group showed that these charge transfer transitions are sensitive to the different scattering geometries[40]. The structure at ~19.0 eV was assigned to the U *6p$_{3/2}$ – 3d$_{5/2}$* transitions [38, 41]. The resolution of the experimental data in Ref. [38] was already good enough to detect an asymmetrical profile of excitations at 0-2eV energy transfer.

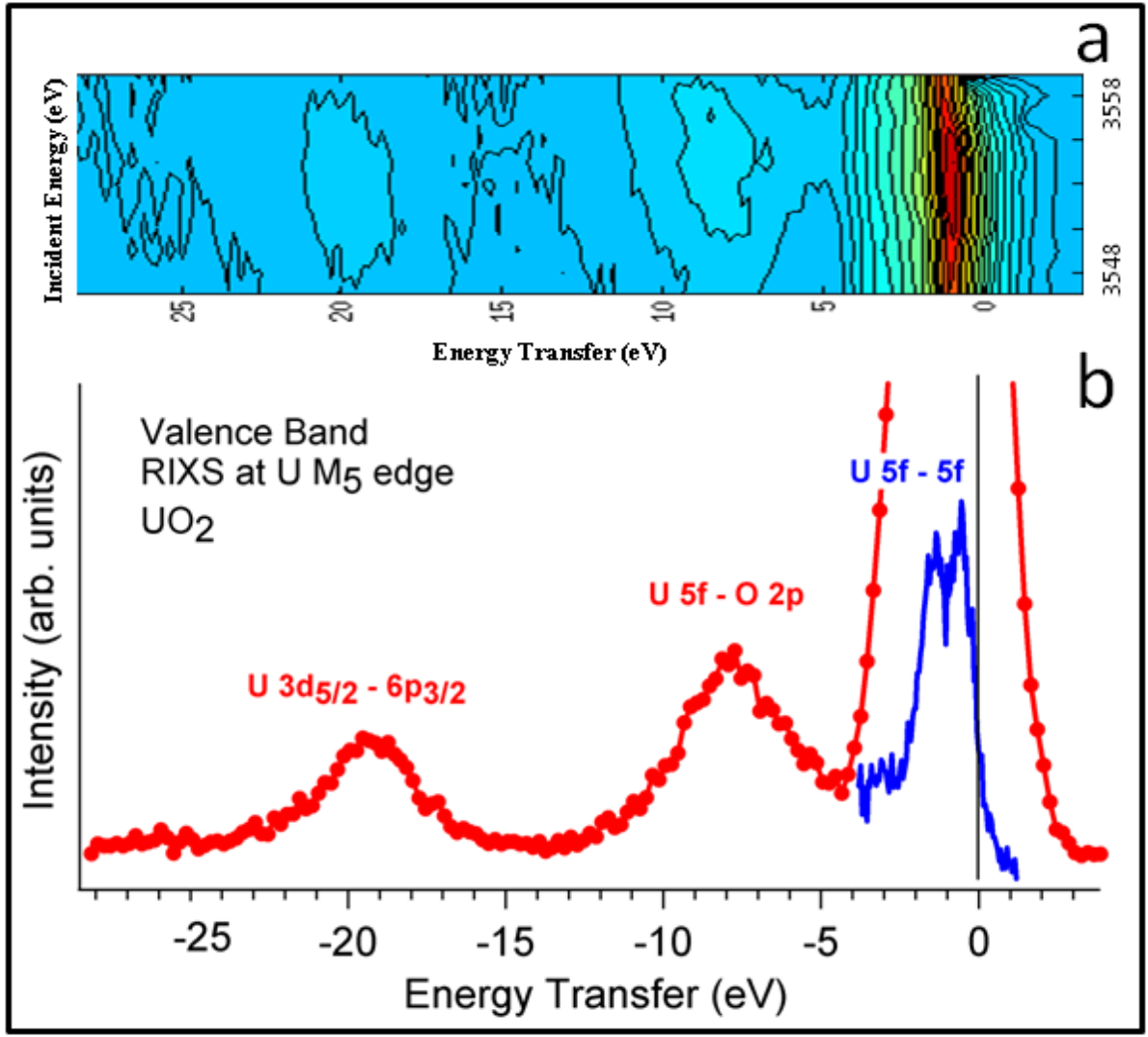

Fig. 10 a) Valence-band RIXS intensities displayed as a contour map with axes corresponding to the incident and transferred energies over the U $M_5$ absorption edge of $UO_2$; b) The U 3d5f RIXS spectrum (red curve) recorded at the incident photon energy of 3552 eV with five crystal-analyzers (see Table 1) and the region of the U 5f-5f excitations of the RIXS spectrum (blue curve) recorded with one crystal-analyzer at the 90° scattering geometry in the horizontal plane.

The observed transitions at ~0-2eV are assigned to the intra-atomic U 5f-5f excitations. The RIXS profile of these excitations is defined by the number of the 5f electrons in the f-shell, in other words by the formal oxidation state of uranium. In turn, the U 5f – ligand 2p charge–transfer excitations depend on the crystal structure of the uranium compound and on the chemical environment of the U atoms. That was shown both experimentally and theoretically for a number of uranium systems [42,43] by core-to-valence RIXS experiments at the U $O_{4,5}$ edge (~100 eV) engaging the U 5d-to-5f core excitations. However, there are still some advantages for using the core-to-valence RIXS at the U $M_{4,5}$ edges, particular for actinide compounds. First, the measurements are more bulk-sensitive technique, which help to avoid confusion with additional states due to surface defects. Second, experiments at the U $M_{4,5}$ edge do not require any high vacuum equipment as in the case of soft X-ray spectroscopy. The technique also allows for easier monitoring of in-situ reactions.

Tender X-ray energy (3.5 – 5.5 keV) can be used to probe the electronic excitations at the other U M thresholds. Fig. 11 shows the HERFD-XANES spectra of $UO_2$ at the U $M_2$ and $M_3$ edges compared with the results of the HERFD-XANES measurements at the U $M_{4,5}$ and U $L_3$ edges. The HERFD spectra at the U $M_2$ edge of $UO_2$ were measured by monitoring the maximum intensity of the $3p_{1/2} - 5d_{3/2}$ transitions (~5079 eV). The HERFD spectra at the U $M_3$ edge of $UO_2$ were measured by monitoring the maximum intensity of the $3p_{3/2} - 4d_{3/2}$ transitions (~ 3522 eV). Following the dipole selection rules, the electrons are excited from the $3p_{1/2}$ and $3p_{3/2}$ levels to the U 6d states at the U $M_2$ and $M_3$ edges, respectively. The distribution of the unoccupied U 6d states is probed as in case of the HERFD measurements at the U $L_3$ edge.

When comparing the HERFD spectra at the $M_3$ versus $L_3$ edge in Fig. 11, three factors need to be considered: the core-hole lifetime broadening of the $4d_{3/2}$ versus $3d_{5/2}$ level, the effects of the interaction of these core holes in the final state of the spectroscopic process with U 6d electrons, and the instrumental resolution. Unfortunately, no improvement in the total spectral broadening was obtained in the measurements at the $M_3$ edge in comparison with those at the $L_3$ edge because somewhat higher instrumental resolution was compensated by somewhat shorter lifetime of the $4d_{3/2}$ core hole (see data about core-hole lifetimes for uranium in Ref. 44). The smaller lifetime broadening (~0.9 eV for U $5d_{3/2}$ core hole, according to Ref. 45) was expected in the measurements at the $M_2$ edge, however, the shape of the obtained HERFD-XANES spectrum turns out to be different from those at the $M_3$ and $L_3$ edges. This can be a result of significant interaction of the 5d core-hole with 5f electrons in the final state of the spectroscopic process which would lead to a redistribution of the 5f states and consequently 6d states due to their admixture to the 5f states.

5. Summary

This work provides a brief overview of the advanced X-ray spectroscopic techniques at the U L and M edges that take advantage of the RIXS detection and allow to improve the resolution of XANES experiments. Core-to-core and core-to-valence RIXS at the U $L_3$ edge provides information on the distribution and crystal-field splitting of the U 6d states which are affected by the crystal structure and the chemical environment of U atoms in the materials. At the same time RIXS measurements at the U $M_{4,5}$ edges provide important information on the localized U 5f states that are largely responsible for the unique properties of U systems.

The time required to record the HERFD spectra that are shown in the present study is less than 10 minutes and few hours are required to record a full core-to-valence RIXS plane. An order-of-magnitude gain can be anticipated by increasing the solid angle and the incident flux. Measurements of the core-to-valence and core-to-core RIXS can thus be extended to demanding sample environments such as high pressure and *in-situ* conditions that take advantage of the large penetration depth of the hard X-ray probe. Such experiments can be performed on very dilute systems that give the opportunity to probe other actinides with higher radioactivity than uranium. In connection with the interpretation of experimental results, we show here that a number of theoretical methods can be used for simulation of the spectra. The combination of theoretical and experimental RIXS data thus provides valuable insight into changes in the density of states, ground state configuration and the f-count of actinide systems.

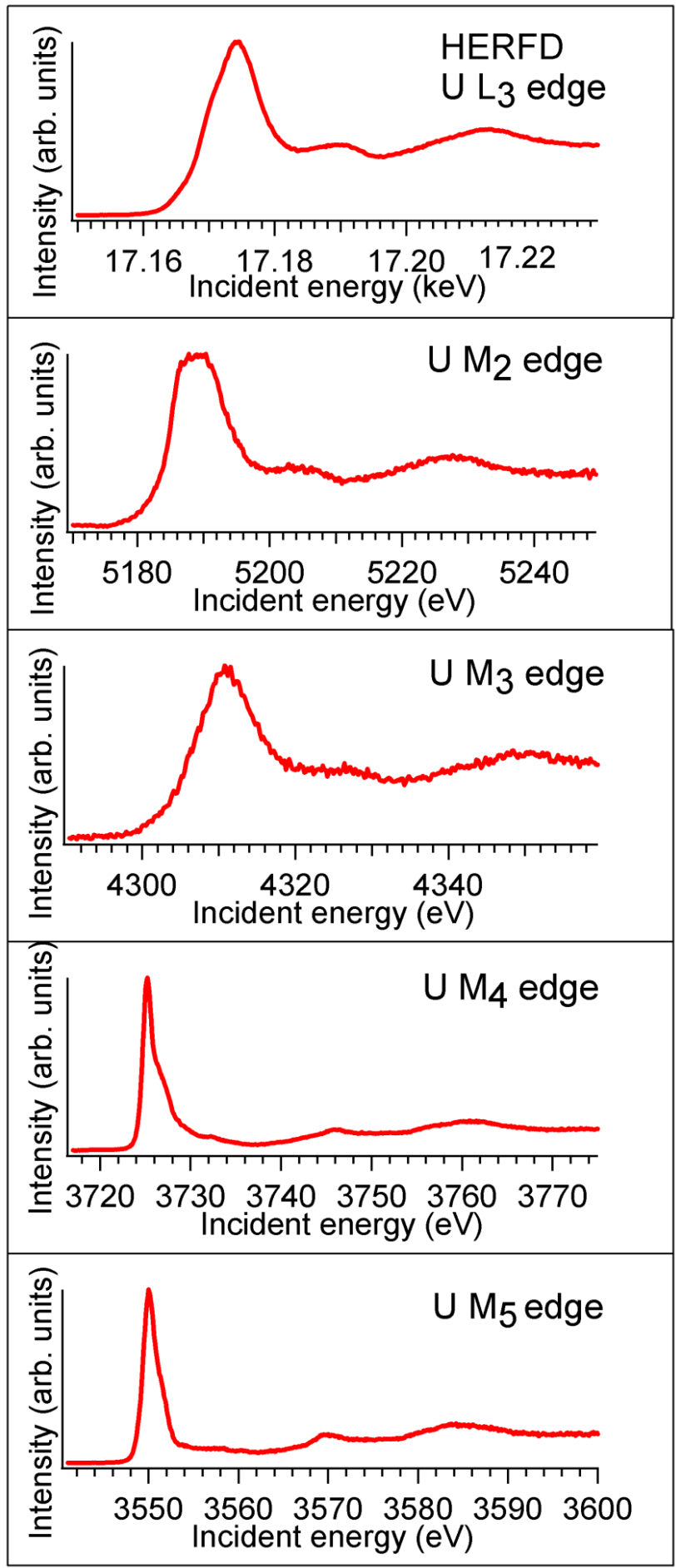

Fig. 11 HERFD-XANES of $UO_2$ at the U $M_2$, $M_3$, $M_4$, $M_5$ and U $L_3$ edges recorded with an X-ray emission spectrometer.

**Acknowledgments**

The authors would like to thank the technical support staff and P. Colomp at the ESRF for the assistance during experiments. K. O. K. would like to thank Y. Kvashnin for performing the LDA+*U* calculations, P. Martin for providing the $UO_2$, $U_3O_8$ and $U_4O_9$ samples, T Behrends for providing the $UO_2(NO_3)_2 \cdot 6H_2O$ sample, and J. Grattage for the English proofreading. K. O. K. is also grateful to A. Bosak for enormous help during a development and implementation of the code for the RIXS calculations. S.M.B acknowledges support from the Swedish Research Council (VR).